\documentclass[prl,twocolumn,aps]{revtex4}
\usepackage{verbatim}
\usepackage{graphicx}
\normalfont
\begin{document}


\title{Suppression of nuclear spin diffusion at a GaAs/AlGaAs interface measured with a single quantum dot nano-probe}


\author{A. E. Nikolaenko$^1$, E. A. Chekhovich$^1$, M. N. Makhonin$^1$, I. W. Drouzas$^1$, A. B. Vankov$^1$, J. Skiba-Szymanska$^1$,  M. S. Skolnick$^1$,  P. Senellart$^2$, A. Lema\^itre$^2$, A. I. Tartakovskii$^1$}

\address{$^{1}$ Department of Physics and Astronomy, University of Sheffield, S3 7RH,UK \\ $^{2}$ Laboratoire de Photonique et de Nanostructures, Route de Nozay, 91460 Marcoussis, France}
\date{\today}

\begin{abstract}

Nuclear spin polarization dynamics are measured in optically pumped individual GaAs/AlGaAs interface quantum dots by detecting the time-dependence of the Overhauser shift in photoluminescence (PL) spectra. Long nuclear polarization decay times of $\approx 1$ minute have been found indicating inefficient nuclear spin diffusion from the GaAs dot into the surrounding AlGaAs matrix in externally applied magnetic field. A spin diffusion coefficient two orders lower than that previously found in bulk GaAs is deduced.  
\\
\end{abstract}

\maketitle


Nuclear spin effects in semiconductors have attracted close attention for several decades. Recently these phenomena came into focus in the field of single electron spin manipulation in semiconductor nano-structures \cite{Petta,Koppens,Atature,GDutt,Reilly}. The hyperfine interaction between the electron and nuclear spins in such semiconductor structures \cite{Overhauser,book} can serve as a powerful tool for controlling spin properties of the localized electron. In particular, polarization of nuclear spins occurring in GaAs-based heterostructures under circularly polarized excitation causes local Overhauser fields up to several Tesla \cite{Gammon1,Gammon2,Bracker,Eble,Lai,Braun1,Maletinsky1,Maletinsky2,Tartakovskii,Makhonin}, leading to marked splittings of the electron spin states, and as a result, to a significant modification of the electron spin coherence and life-time \cite{Petta,Koppens,Atature,GDutt,Reilly,Khaetskii,Erlingsson,Merkulov,Braun2,Reilly}. 
In such conditions the nuclear spin dynamics on the dot sets the time-scales for operations on the electron spin in a controlled magnetic environment. The nuclear spin dynamics are influenced mainly by the hyperfine interaction with electron spins and also the dipole-dipole coupling between nuclei. In particular, the latter process leads to so-called nuclear spin diffusion \cite{Bloembergen,Paget}. 

In this work we introduce a new approach to measurements of the nuclear spin diffusion in GaAs/AlGaAs heterostructures, the type of structures widely employed in electron \cite{Petta,Koppens,Reilly} and nuclear spin \cite{Sanada,Yusa} coherent control experiments and quantum Hall effect measurements \cite{Smet}. The Overhauser shift (OHS) of a single electron spin state in individual QDs, acts as an accurate local probe, which provides a direct and quantitative measure of the degree of polarization of 10$^4$-10$^5$ nuclear spins \cite{Gammon1,Gammon2,Bracker,GDutt}. In this work we use such a nano-probe positioned in a mono-layer fluctuation quantum dot to monitor the nuclear spin diffusion at the interface of a GaAs quantum well (QWs) and an AlGaAs barrier. This provides important information on the nuclear spin dynamics on the nano-scale, removing the effects of sample inhomogeneities typical for macroscopic structures.

We find very slow nuclear polarization decay with characteristic times of $\approx$1 min in an individual interface QD in the regime of high magnetic fields.  By employing a 3D diffusion model, we obtain very good fits to the decay kinetics using a surprisingly low diffusion coefficient of $\approx2\cdot10^{-15}$cm$^2$/s, two orders of magnitude lower than was found for bulk GaAs by Paget \cite{Paget}. We also report nuclear polarization rise times in the range of 0.5-5 s dependent on optical pumping power, polarization of excitation and the magnitude of the external magnetic field. 

The sample investigated contains a nominally 13-monolayer GaAs quantum well (QW) embedded in Al$_{0.33}$Ga$_{0.67}$As barriers (see growth detail in Ref.\cite{Peter}). Interface QDs are formed naturally by 1 monolayer QW width fluctuations. With the lateral dimensions on the order of 10-100 nm these potential fluctuations result in up to 15 meV change in the exciton energy sufficient for zero dimensional exciton localization at low temperatures \cite{Gammon1,Gammon2,Bracker,Peter}. A 40/10/90 nm SiO$_{2}$/Ti/Al shadow mask was deposited on the sample surface, with 800 nm diameter apertures opened for optical access to individual dots. The nuclear spin dynamics were measured at a temperature $T$=4.2 K with magnetic field of several Tesla applied in the Faraday geometry \cite{lowfield}. PL  was detected with a double spectrometer and a CCD. A laser at 670 nm was employed to generate electrons and holes in the QW states $\approx$130 meV above the QDs emission lines: absorption in the AlGaAs barriers at this wavelength is negligible. 

\begin{figure}
\centering
\includegraphics[width=8cm]{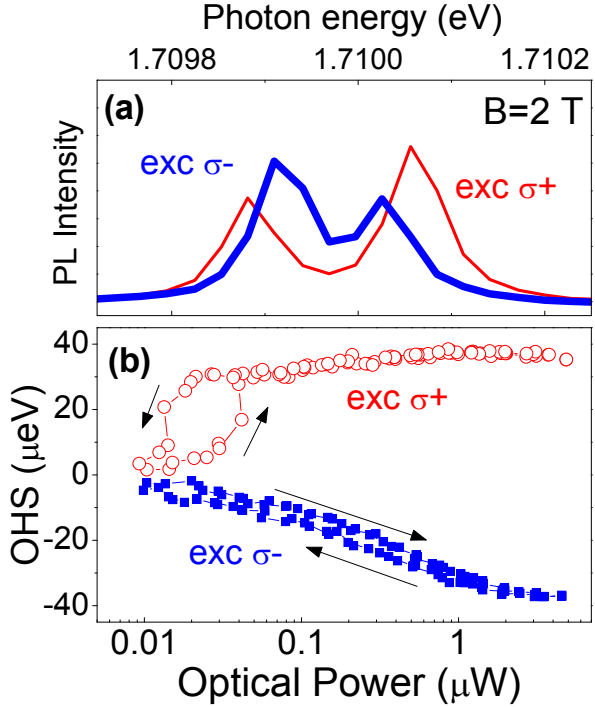}
\caption{(color online). (a) Thin (thick) line shows a low-temperature PL spectrum of a single GaAs/AlGaAs dot excited with $\sigma^+$($\sigma^-$) pump in external magnetic field of 2T. (b) Power dependences of the exciton Zeeman splitting for excitation with $\sigma^+$ (circles) and $\sigma^-$ (squares) polarizations. Arrows show the direction in which the power was scanned.}
\label{fig1}
\end{figure}

Fig.1a shows unpolarized PL spectra measured for an individual GaAs/AlGaAs dot in an external magnetic field $B_{z}$=2 T for excitation with $\sigma^+$ and $\sigma^-$ circularly polarized light (red and blue lines, respectively). These peaks correspond to recombination of a neutral exciton localized in the dot. The observed doublet is due to the exciton Zeeman splitting modified by the Overhauser shift of the electron spin states. The collective effect of the hyperfine interaction of the nuclear spins in the QD with the photo-generated electron can be treated as an additional magnetic
field $B_N$ \cite{Overhauser,book}, that builds up dynamically under circularly polarized optical excitation \cite{book}. $B_N$ will act together with the external magnetic field $B_z$ resulting in a spectral doublet (as seen in Fig.1a) with a splitting $\Delta E(\sigma^{\pm})$=$\mu_B[|g_h| B_z-|g_e|(B_z\mp B_N)]$ dependent on the polarization of excitation. Here $g_{e(h)}$ is the electron (hole) g-factor and  $\mu_B$ is the Bohr magneton. In the rest of this work we use $\Delta E$ to probe nuclear spin polarization on the dot.

Fig.1b shows the dependences of the OHS on excitation power for $\sigma^+$ and $\sigma^-$ polarized excitation. Qualitatively different dependences are observed in the two cases: a strong threshold-like $B_N$ switching and large hysteresis loop is observed for $\sigma^+$ excitation, whereas a smooth curve with very weak hysteresis is measured in the $\sigma^-$ case. The marked difference originates from feedback of the optically induced Overhauser field on the dot on the electron-to-nuclei spin transfer efficiency occurring due to the dependence of the electron Zeeman splitting (the major energy cost of the electron-nuclear spin flip-flop) on both $B_z$ and $B_N$ \cite{Eble,Maletinsky1,Braun1,Tartakovskii,book}. The additional electron spin state splitting produced by $B_N$ either enhances the spin transfer when the electron Zeeman splitting $E_{eZ}(\sigma^{\pm})=\mu_Bg_e(B_z\mp B_N)$ is reduced (for $\sigma^+$) or lead to inefficient spin pumping when $E_{eZ}$ is increased (for $\sigma^-$). The strong feedback in the case of $\sigma^+$ excitation leads to the clear bistable behavior in Fig.1b \cite{Maletinsky1,Braun2,Tartakovskii,Makhonin}.  

Note, that saturation of the OHS at an absolute value of 38 $\mu$eV is observed at high power for both polarizations. The maximum OHS of $\approx$38$\mu$eV corresponds to a nuclear polarization degree of $\approx29\%$, as deduced from the maximum OHS in a fully polarized dot of $\delta_{n100\%}$=$I_{Ga}A_{Ga}+I_{As}A_{As}$=132 $\mu$eV. Here $A_{Ga}$=42 $\mu$eV and $A_{As}$=46 $\mu$eV are the hyperfine constants and $I_{Ga}$=$I_{As}$=3/2 are the spins of Ga and As nuclei \cite{Paget}.

\begin{figure}
\centering
\includegraphics[width=8cm]{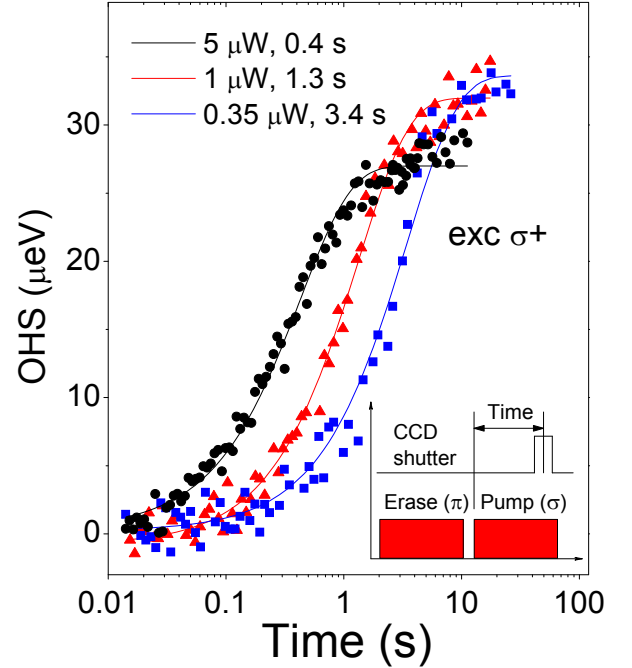}
\caption{(color online). Nuclear polarization built-up dynamics in a GaAs/AlGaAs interface QD measured in experiment schematically showed in the diagram. The curves are obtained for the excitation power of 5 $\mu$W (circles), 1 $\mu$W (triangles) and 0.35 $\mu$W (squares) of $\sigma^+$ pump. The times shown are the results of a single exponent fitting (solid curves) of the experimental data.}
\label{fig2}
\end{figure}

In this work the nuclear spin life-time is measured employing an all-optical pump-probe method, where the effect of the strong pump is tested after a time-delay with a weak probe. In order to identify the appropriate length and optical power of the probe so that it does not perturb the system, the nuclear polarization rise time is
firstly measured. This is reported in Fig.2 for a single optically pumped GaAs/AlGaAs dot in an external magnetic field $B_{ext}$ of 2 T. The optical pulse sequence for this measurement is schematically shown in the inset of Fig.2. The pulses from two lasers are prepared by fast mechanical shutters with a rise time $<$2 ms. The first laser pulse (denoted as 'erase') is 10 s long and has linear polarization. This pulse is required for destruction of any nuclear polarization in the dot and surrounding QW remaining from the previous measurement cycle.  Nuclear polarization is then pumped by the second circularly polarized 'pump' pulse of variable length. A mechanical shutter placed in the PL detection path allows PL signal acquisition in a narrow time window at the end of the pump pulse. The time resolution of the dynamics measurement is determined by the time this shutter is open. For the shortest delay times the resolution was as low as 5 ms. In order to improve signal to noise ratio in the PL spectra measured for each delay, sequences of the erase and pump laser pulses and the PL acquisition were repeated several times for each spectrum measured. The degree of nuclear polarization for each delay was determined from the modification of the exciton Zeeman splitting due to the OHS.

We find a strong dependence of the nuclear spin rise time on the pumping power.
Fig.2 shows evolution of the OHS with time under the $\sigma^+$ polarized pumping  measured with different optical excitation powers. The measured dynamics curves in Fig.2 can be well approximated with exponential fits with rise times of 0.4, 1.3, and 3.4 s for excitation power of 5, 1 and 0.35 $\mu$W, respectively \cite{Belhadj}. These findings can be explained by the power-dependent supply rate of the spin polarized electrons to the dot: faster nuclear spin pumping occurs when the spin supply rate is increased at a higher excitation power \cite{Broun1,Tartakovskii}.

In order to investigate the nuclear polarization decay the pulse sequence was modified into a 'pump-probe' configuration schematically shown in the inset of Fig.3. Nuclear polarization is pumped by a 10 s long circularly polarized pump pulse. Then a short linearly polarized pulse is applied at a variable delay after the end of the pump pulse to probe the exciton Zeeman splitting. Based on the measured nuclear polarization built-up dynamics shown in Fig.2 we set the length of the probe pulse to 100 ms to minimize its effect on the nuclear polarization in the dot. In the time period between the two pulses the sample is kept in the dark. The CCD shutter is open during the probe pulse only. In order to improve signal to noise ratio several pump-probe measurement cycles are performed with no dark time between the cycles as the dot is polarized to saturation by the long pump pulse independently on the initial polarization.

Fig.3 shows the nuclear spin decay dynamics measured at $B_{ext}$=2T. The exciton Zeeman splitting is plotted as a function of the 'dark' delay time between the pump and probe pulses. As seen from the figure decay for both polarizations of the pump occurs with a characteristic time of $\approx$1 min.

The most likely mechanisms for the nuclear spin depolarization ''detected'' by the QD nano-probe is nuclear spin diffusion into the unpolarized barrier. Two other mechanisms that could contribute are depolarization due to interaction with an electron gas in the QW and phonon-assisted spin-lattice relaxation. The relaxation via interaction with residual electrons can be excluded since the sample is nominally undoped and no evidence for charging in either the QW or dots is found in PL. The quadrupolar relaxation via phonon-assisted processes is rather weak for temperatures below 30K \cite{Mcneil} leading to characteristic decay times of the order of 1000 s in GaAs. On the other hand, the nuclear polarization in the extremely small volume of the dot is very sensitive to the spin ''leakage'' into the surrounding bulk. In what follows we will focus on this decay mechanism.

Decay of the nuclear polarization $S_N$ due to diffusion can be described with a standard 3D diffusion equation 
\begin{equation}
dS_N(r,t)/dt=D_{QD}\Delta S_N(r,t),
\label{diffusion}
\end{equation}
where $D_{QD}$ is the nuclear spin diffusion coefficient.  In our modeling we approximate the dot shape with a 5 nm high disk with a 20 nm diameter and assume that the nuclear spin is pumped optically inside the dot volume, whereas the material around the dot is not polarized directly by optical excitation. This description is an approximation of the actual nuclear spin pumping process. In a more complex model, spatially non-uniform optical pumping of the nuclear spin in the whole two-dimensional sheet of the QW should be taken into account. From our calculations we find, however, that the low aspect ratio of the model dot effectively leads to a one-dimensional spin diffusion process, rather insensitive to the nuclear polarization in adjacent parts of the well. In what follows we assume that the nuclei outside the dot are polarized only via spin diffusion from the dot.

Initially, at the beginning of the pumping pulse, the dot and surrounding material are not polarized. We first calculate the distribution of the nuclear spin due to the diffusion from the dot in the first 10 seconds during the optical pulse. For this we assume an instantaneous increase of the polarization inside the dot, which is kept constant for the duration of the pumping pulse. Using this distribution as the initial condition (the high polarization in the dot and the decreasing polarization away from the dot in the bulk), we then simulate the time-evolution of the nuclear spin polarization on the dot in the dark by allowing the polarization inside the dot to decay with time. 

The only fitting parameter that we vary in our model is the diffusion coefficient $D_{QD}$.  Solid curves in Fig.3 show the results obtained for three different diffusion coefficients. The thin black line shows the decay with $D_{QD}=10^{-13}$cm$^2$/s found by Paget in Ref.\cite{Paget} for bulk GaAs. Our calculations predict that for this value of diffusion coefficient the spin polarization on the dot falls to $30\%$ level after 5 second, about 12 times faster than found in our experiment on GaAs/AlGaAs dots. The thick black line providing an excellent fit to our data shows results for a very low magnitude of $D_{QD}=2\cdot10^{-15}$cm$^2$/s. This is about 20 times smaller than that we recently found for InGaAs/GaAs dots \cite{Makhonin} and 50 times smaller than in bulk GaAs \cite{Paget}. To exclude a scenario where the volume around the dot is polarized by a very fast diffusion during the 10 s pumping time, which then would prevent further spin leakage from the dot,  we also show results for a high magnitude of $D_{QD}=10^{-12}$cm$^2$/s (gray line in Fig.3). As can be seen, an even faster decay of nuclear spin on the dot is observed in that case with the decay time of 1 s. We thus conclude that in the regime of high magnetic field in this work the nuclear spin diffusion in the studied GaAs/Al$_{0.3}$Ga$_{0.7}$As heterostructure is strongly suppressed \cite{lowfield}.

In magnetic field due to a non-zero nuclear Zeeman splitting the spin diffusion can only occur via transfer of spin between the like isotope species, since different isotopes exhibit differing energy splittings. Since the optical pumping in our experiment occurs only in the GaAs QW, we suggest that the slowing of the nuclear
diffusion occurs in AlGaAs or at the GaAs/AlGaAs interface. There the dipole-dipole interaction between the Ga spins (the mechanism underlying the diffusion) is weakened due to a larger distance between the like nuclei. The considerably more efficient spin leakage in the InGaAs/GaAs system probably occurs due to the efficient diffusion of the spins of Ga and As nuclei into GaAs around the dot.

\begin{figure}
\centering
\includegraphics[width=8cm]{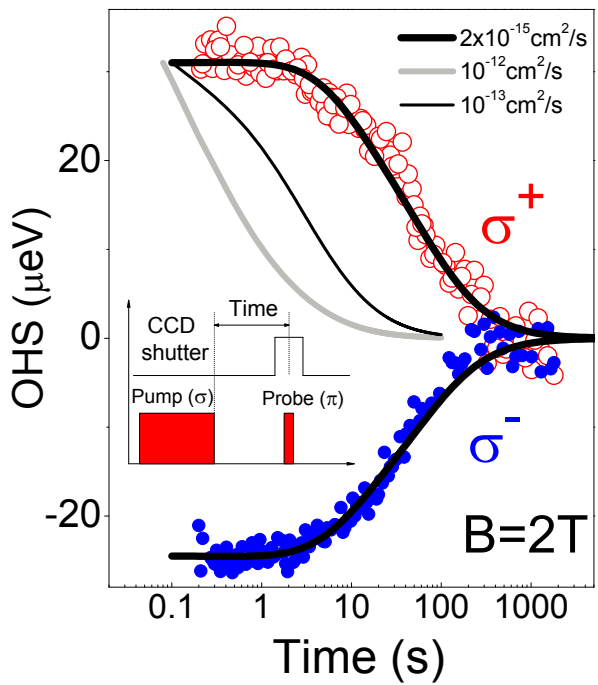}
\caption{(color online). Nuclear polarization decay curves measured for GaAs/AlGaAs interface QDs for two polarizations of the pump pulse in external magnetic field of 2T (circles). Lines show polarization decay curves calculated using Eq.\ref{diffusion} with   $D_{QD}=2\cdot10^{-15}$ cm$^2$/s (thik black), $10^{-13}$ cm$^2$/s (thin black) and $10^{-12}$ cm$^2$/s (gray).} 
\label{fig3}
\end{figure}

In summary, we have measured nuclear polarization dynamics in a single interface GaAs/Al$_{0.3}$Ga$_{0.7}$As QDs under non-resonant circularly polarized optical excitation. The rise time is found to be sensitive to the excitation power and external magnetic field and varies in the range of $\approx 0.4\div5$ seconds. A very slow rate of the nuclear polarization decay has been measured with a characteristic depolarization time exceeding 1 min in magnetic field of several Tesla. This indicates suppression of the nuclear spin diffusion from the GaAs dot into the AlGaAs barrier. We find an almost 2 orders of magnitude smaller diffusion coefficient compared to bulk GaAs and InGaAs dots reported recently. This observation can be  explained by the increased distance at the interface of the dot and in the barrier between the like nuclear species contributing to the flip-flop-like diffusion process in the external magnetic field. The increased separation leads to a notable weakening of the dipole-dipole interaction and, consequently, to a slower nuclear spin decay in a GaAs/AlGaAs dot.  

This work has been supported by  EPSRC grants EP/C54563X/1 and EP/C545648/1, Programme Grant GR/S76076, Programme Grant EP/G601642/1, the EPSRC IRC for Quantum Information Processing, and by the Royal Society.


\begin{thebibliography}{}

\bibitem{Petta} J. R. Petta {\it et al.}, Science  {\bf 309}, 2180 (2005).
\bibitem {Koppens} F. H. L. Koppens {\it et al.}, Science {\bf 309}, 1346, (2005).
\bibitem{Atature}  M. Atatüre, J. Dreiser, A. Badolato, A. Högele, K. Karrai, A. Imamoglu, Science{\bf 312}, 551 (2006).
\bibitem{GDutt} M. V. Gurudev Dutt, J. Cheng, Bo Li, X. Xu, X. Li, P. R. Berman, D. G. Steel, A. S. Bracker, D. Gammon, S. E. Economou, Ren-Bao Liu, L. J. Sham, Phys. Rev. Lett. {\bf 94}, 227403 (2005).
\bibitem{Reilly} D. J. Reilly, J. M. Taylor, J. R. Petta, C. M. Marcus, M. P. Hanson, A. C. Gossard, Science {\bf 321}, 817 (2008).
\bibitem{Overhauser} A. W. Overhauser, Phys. Rev.{\bf 92}, 411 (1953).
\bibitem{book} Optical Orientation, edited by F. Meier and B. P. Zakarchenya (Elsevier, New York, 1984).
\bibitem{Gammon1} D. Gammon, S. W. Brown, E. S. Snow, T. A. Kennedy, D. S. Katzer, and D. Park, Science {\bf 277}, 85 (1997).
\bibitem{Gammon2} D. Gammon, A. L. Efros, T. A. Kennedy, M. Rosen, D. S. Katzer, D. Park, S. W. Brown, V. L. Korenev, and I. A. Merkulov, Phys. Rev. Lett. {\bf 86}, 5176 (2001).
\bibitem{Bracker} A. S. Bracker, E. A. Stinaff, D. Gammon, M. E. Ware, J. G. Tischler, A. Shabaev, Al. L. Efros, D. Park, D. Gershoni, V. L. Korenev, and I. A. Merkulov, Phys. Rev. Lett. {\bf 94}, 047402 (2005).
\bibitem{Eble} B. Eble, O. Krebs, A. Lemaitre, K. Kowalik, A. Kudelski, P. Voisin, B. Urbaszek, X. Marie, and T. Amand, Phys. Rev. B {\bf 74}, 081306(R) (2006).
\bibitem{Lai} C. W. Lai, P. Maletinsky, A. Badolato, and A. Imamoglu, Phys. Rev. Lett. {\bf 96}, 167403 (2006).
\bibitem{Braun1} P.-F. Braun, B. Urbaszek, T. Amand, X. Marie, O. Krebs, B. Eble, A. Lemaitre, and P. Voisin, Phys. Rev. B {\bf 74}, 245306 (2006).
\bibitem{Maletinsky1} P. Maletinsky, C. W. Lai, A. Badolato, and A. Imamoglu, Phys. Rev. B {\bf 75}, 035409 (2007).
\bibitem{Tartakovskii} A. I. Tartakovskii, T. Wright, A. Russell, V. I. Fal'ko, A. B. Van'kov, J. Skiba-Szymanska, I. Drouzas, R. S. Kolodka, M. S. Skolnick, P.W. Fry, A. Tahraoui, H.-Y. Liu, and M. Hopkinson, Phys. Rev. Lett. {\bf 98}, 026806 (2007).
\bibitem{Maletinsky2} P. Maletinsky, A. Badolato, and A. Imamoglu, Phys. Rev. Lett. {\bf 99}, 056804 (2007).
\bibitem{Makhonin} M. N. Makhonin, A. I. Tartakovskii, A. B. Van'kov, I. Drouzas, T. Wright, J. Skiba-Szymanska, A. Russell, V. I. Fal'ko, M. S. Skolnick, H.-Y. Liu, and M. Hopkinson, Phys. Rev. B {\bf 77}, 125307 (2008).
\bibitem{Khaetskii} A. V. Khaetskii, D. Loss, L. Glazman, Phys. Rev. Lett.{\bf  88}, 186802 (2002).
\bibitem{Erlingsson} S. I. Erlingsson, Y. V. Nazarov, V. I. Fal'ko, Phys. Rev.  {\bf B 64}, 195306 (2001).
\bibitem{Merkulov} I. A. Merkulov, Al. L. Efros, and M. Rosen, Phys. Rev. B {\bf 65}, 205309 (2002).
\bibitem{Braun2} P.-F. Braun, X. Marie, L. Lombez, B. Urbaszek, T. Amand, P. Renucci, V. K. Kalevich, K. V. Kavokin, O. Krebs, P. Voisin, and Y. Masumoto, Phys. Rev. Lett. {\bf 94}, 116601 (2005).
\bibitem{Bloembergen} N. Bloembergen, Physica (Utrecht) {\bf 15}, 386 (1949).
\bibitem{Sanada} H. Sanada, Y. Kondo, S. Matsuzaka, K. Morita, C.Y. Hu, Y. Ohno, H. Ohno, Phys. Rev. Lett. {\bf  96}, 067602 (2006).
\bibitem{Yusa} G. Yusa, K. Muraki, K. Takashina, K. Hashimoto, Y. Hirayama, Nature {\bf 434} 1001 (2005).
\bibitem{Smet} J. H. Smet, R. A. Deutschmann, F. Ertl, W. Wegscheider, G. Abstreiter, K. von Klitzing, Nature {\bf 415} 281 (2002).
\bibitem{Paget} D. Paget, Phys. Rev. B {\bf 25}, 4444 (1982).
\bibitem{Peter} E. Peter {\it et al}, Phys. Rev. Lett. {\bf 95}, 067401 (2005).
\bibitem{lowfield} Negligible nuclear spin pumping is observed in magnetic fields below 20 mT, indicating efficient depolarization due to dipole-dipole interaction. 
\bibitem{Belhadj} Similar rise times have been observed in GaAs/AlGaAs dots grown by droplet epitaxy in T. Belhadj {\it et al}, Phys. Rev. {\bf B 78}, 205325 (2008).
\bibitem{Mcneil} J. A. McNeil, W. G. Clark, Phys. Rev. {\bf B 13}, 4705 (1976).

\end{thebibliography}
\end{document}